\documentclass[doublecol,linenumbers]{epl2} % for 2 columns style with line numbers
\title{The Momentum Distribution of Liquid $^3$He}
\shorttitle{The Momentum Distribution of Liquid $^3$He}

\author{M.S. Bryan\inst{1} \and T.R. Prisk\inst{2} \and R.T. Azuah\inst{2,3} \and W.G. Stirling\inst{4} \and P.E. Sokol\inst{1}}
\shortauthor{M.S. Bryan \etal}

\institute{                    
  \inst{1} Department of Physics, Indiana University, Bloomington, IN 47408, USA\\
  \inst{2} Center for Neutron Research, National Institute of Standards and Technology, Gaithersburg, MD 20899-6100, USA\\
  \inst{3} Department of Materials Science and Engineering, University of Maryland, College Park, MD 20742-2115, USA\\
  \inst{4} Institute Laue-Langevin, Grenoble, France
}

\abstract{We present high-resolution neutron Compton scattering measurements of liquid $^3$He below its renormalized Fermi temperature.  Theoretical predictions are in excellent agreement with the experimental data when instrumental resolution and final state effects are accounted for.  Our results resolve the long-standing inconsistency between theoretical and experimental estimates of the average atomic kinetic energy.}
\begin{document}

\maketitle

\section{Introduction}
Liquid $^3$He is a system of fundamental importance to contemporary physics because it is a prototypical example of a strongly interacting fermion system\cite{Baym, Pines}.  It provides a benchmark for testing the reliability of current many-body techniques, such as variational wavefunctions\cite{Holzmann, Manousakis} and quantum Monte Carlo calculations\cite{Casulleras, Ceperley, Panoff, Mazzanti, Moroni1}.  Fermi statistics plays a dominant role in determining the microscopic dynamics of normal liquid $^3$He at both collective and single-particle levels.  The elementary excitations of liquid $^3$He consist of particle-hole quasi-particles and zero sound\cite{Glyde, Fak, Albergamo}.  The atomic momentum distribution $n(k)$ is believed to exhibit a Fermi surface discontinuity followed at higher momenta by an exponential tail.  The single-particle dynamics of liquid $^3$He shares these universal features of normal Fermi liquids with electronic and nuclear systems, despite the large differences in length and energy scales involved\cite{CooperBook, Cooper, Benhar, SilverSokol}.

Variational wavefunctions and Diffusion Monte Carlo (DMC) methods have been applied to determining the momentum distribution $n(k)$ of liquid $^3$He\cite{Manousakis, Moroni1, Mazzanti}.  These theories predict that the Fermi surface discontinuity $Z$ has a size of $\approx 0.25$ and the average kinetic energy $\langle E_K\rangle$ is about 12 K to 13 K, under saturated vapor pressure.  A number of experimenters have performed neutron Compton scattering studies of liquid $^3$He to test this Fermi liquid picture of $^3$He\cite{Sokol, Mook, Azuah}.  No direct observation of the Fermi surface discontinuity has resulted from these measurements.  Moreover, these authors report values for the average atomic kinetic energy $\langle E_K\rangle$ in the range of 8 K to 10 K, in serious disagreement with theoretical predictions.  It was suggested that the experimental estimates of $\langle E_K\rangle$ may be incorrect due to the fact that there is a significant contribution to $\langle E_K \rangle$ from the tails of the data, the most poorly known part of the scattering\cite{CarlsonR, SokolR}.  This long-standing inconsistency casts doubt upon the ability of modern many-body methods to describe this benchmark fermion system.

In this paper, we present a high-resolution neutron Compton scattering study of liquid $^3$He below its renormalized Fermi temperature $T_F^*$.   A preliminary report of the experiment has already appeared\cite{Dimeo}.  We demonstrate that the neutron Compton profile $J(Y, Q)$ and average kinetic energy $\langle E_K\rangle$ are fully consistent with many-body predictions.  There is excellent agreement between the predicted and observed lineshape of $J(Y, Q)$ when instrumental resolution and final state effects are taken into account.  The average kinetic energy $\langle E_K\rangle$, extracted from the scattering data by means of a model fit, is in agreement with theoretical predictions for pressures of 0 bar to 15 bar.  Therefore, the Fermi liquid picture of normal fluid $^3$He is fully consistent with our scattering data.

\section{Theoretical Predictions}
\begin{figure}
		\includegraphics[width=\linewidth]{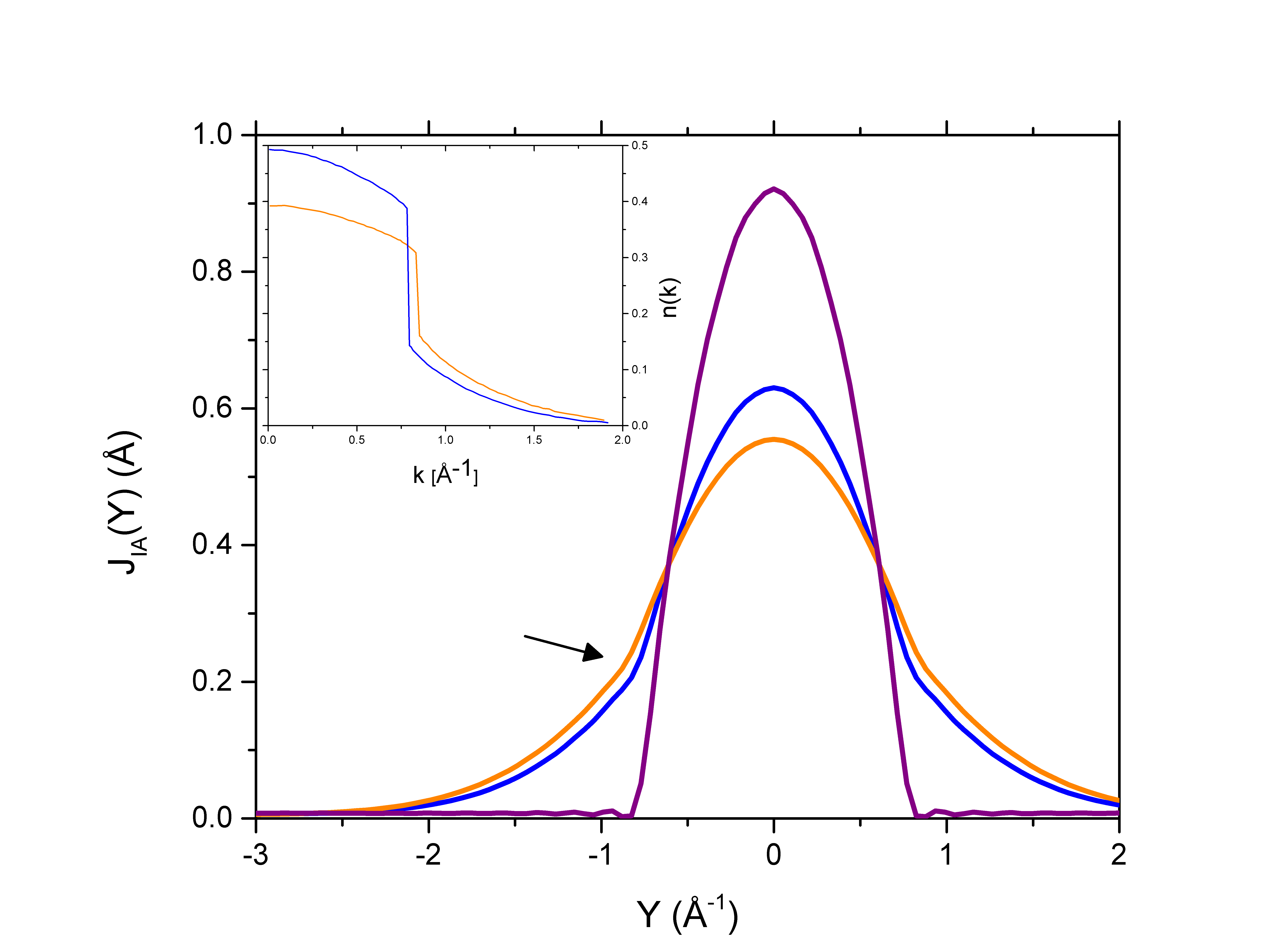}
		\caption{Expected IA-scattering $J_{IA}(Y)$ based on DMC predictions described in Ref\cite{Moroni1}.  Main panel: liquid $^3$He at number densities $0.01635 \textrm{ \AA}^{-3}$ (blue) and $0.01946 \textrm{ \AA}^{-3}$ (orange); ideal Fermi gas at number density $0.01635 \textrm{ \AA}^{-3}$ (purple).  The arrow points to the Fermi surface kink at $Y = -k_F$. Inset: corresponding momentum distributions $n(k)$.}
		\label{fig:IA}
	\end{figure}
	
Here we consider the DMC calculations of Moroni \emph{et al}\cite{Moroni1} and their implications for the neutron scattering law in the Impulse Approximation (IA) limit.  The inset of Figure 1 illustrates the momentum distribution $n(k)$ of liquid $^3$He at number densities of $0.01635 \textrm{ \AA}^{-3}$ and $0.01946 \textrm{ \AA}^{-3}$, corresponding to pressures of 0 bar and 10 bar.  Our experimental data, discussed below, also covers these pressures.  The momentum distribution $n(k)$ exhibits a sharp discontinuity at the Fermi wavevector $k_F$ and an exponential drop-off for $k > k_F$.  As the pressure is raised, the size $Z$ of the discontinuity decreases from 0.24 to 0.14 and more $^3$He atoms are promoted from the ``Fermi sea" into the exponential tails.   The average kinetic energy $\langle E_K\rangle$ is increased from 12.0 K to 15.2 K. 

Neutron Compton scattering experiments use epithermal neutrons to reach momentum and energy transfers that are much larger than the characteristic energies of the material under study\cite{SilverSokol, GlydeBook, Andreani}.  According to the IA, a high-energy incident neutron delivers an impulsive blow to a single helium atom in the sample, transferring a sufficiently large amount of momentum and kinetic energy to the target atom so that it recoils freely from the impact.  In this limit, the neutron Compton profile $J(Y, Q)$ is given directly by an integral transform of the momentum distribution $n(k)$.
\begin{eqnarray}
J_{IA}(Y) &=& 2\pi\int_{|Y|}^\infty kn(k)dk.
\label{eq:JIA}
\end{eqnarray}
The main panel of Figure \ref{fig:IA} compares the IA-scattering law $J_{IA}(Y)$ for three different cases.  The scattering is exactly parabolic in the ideal Fermi gas when $|Y| \le k_F$ and zero otherwise.  In liquid $^3$He $J_{IA}(Y)$ is approximately, but not exactly, parabolic when $-k_F \le Y \le +k_F$.  However, unlike the ideal Fermi gas, strong interactions in the liquid promote particles to momenta $k$ above the Fermi surface $k_F$.   Under the integral transform in Equation \ref{eq:JIA}, the Fermi surface discontinuity in $n(k)$ becomes a \emph{kink} in $J_{IA}(Y)$ at $Y = \pm k_F$.

The IA limit is not reached in neutron scattering studies of the helium liquids because the interatomic potential has steeply repulsive core.  Final state effects (FSE) are deviations from the IA that occur because the recoiling helium atom may collide with the hard cores of its neighbors.  FSE smear or wash out sharp features in the neutron Compton profile.  They are expressed as a broadening function $R(Y, Q)$:
\begin{equation}
J_{FS}(Y, Q) = \int_{-\infty}^{+\infty} J_{IA}(Y^\prime)R(Y - Y^\prime, Q)dY^\prime.
\end{equation}
In this paper, we adopt the model FSE function $R(Y, Q)$ from Hard Core Perturbation Theory (HCPT)\cite{Silver1, Silver2, Sosnick, Snow}.  The most detailed HCPT calculations have been performed for liquid $^4$He, although the theory is also applicable to liquid $^3$He.  To obtain an appropriate model FSE function $R(Y, Q)$ of liquid $^3$He we have applied the approximate density-scaling property of HCPT:
\begin{equation}
\frac{Y_1}{Y_2} = \frac{R(Y_2, Q)}{R(Y_1, Q)} = \frac{n_1}{n_2}.
\end{equation}
Here $n_1$ and $n_2$ refer to two different number densities of liquid helium.
\begin{figure}
		\includegraphics[width=\linewidth]{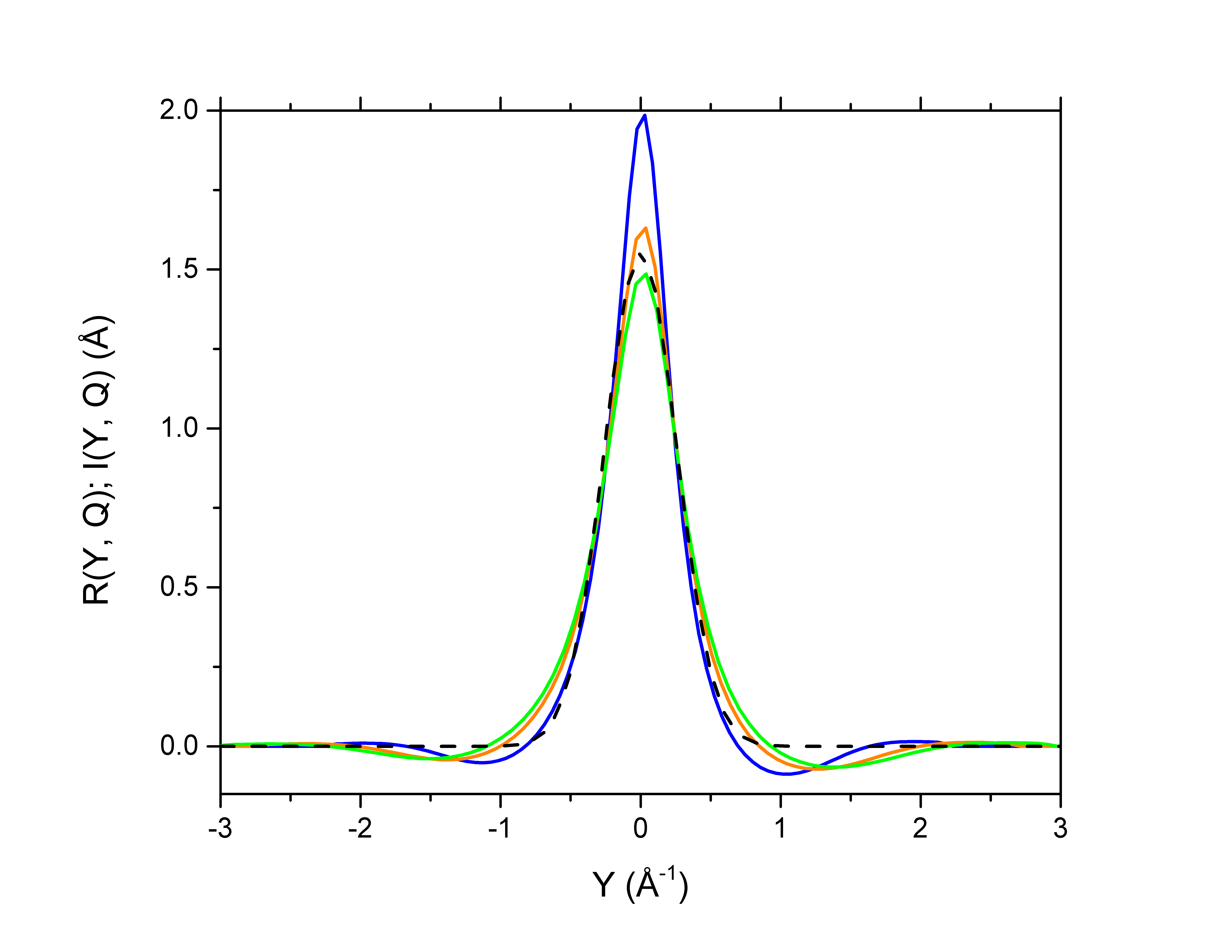}
		\caption{Corrections to the IA: FSE broadening function $R(Y, Q)$ for number densities $0.01635 \textrm{ \AA}^{-3}$ (blue), $0.01946 \textrm{ \AA}^{-3}$ (orange), and $0.0216 \textrm{ \AA}^{-3}$ (green); the effective instrumental resolution $I(Y, Q)$ (dashed black line).}
		\label{fig:FSE}
	\end{figure}

Figure \ref{fig:FSE} plots the final state effect function $R(Y, Q)$ for three different number densities.  $R(Y, Q)$ consists of a central peak and damped oscillatory tails.  The width of the central peak increases at higher densities, and the damped oscillatory tails ensure that the second moment of $R(Y, Q)$ vanishes.  We note that the HCPT theory is very similar in shape and width to the model proposed by Mazzanti \emph{et al}\cite{Mazzanti}.

\section{Experimental Approach}
We carried out a neutron Compton scattering study of liquid $^3$He using the MARI spectrometer at Rutherford Appleton Laboratory.  For this study, we used the A-chopper package to obtain a nominal incident energy $E_i = 800$ meV and an elastic energy resolution $\delta E/E_i \approx 2$ \%.  The neutron Compton profile $J(Y, Q)$ was determined at $Q = 27.5 \textrm{ \AA}^{-1}$, corresponding to a mean scattering angle $\phi = 125^\circ$.
  
The large absorption cross section of $^3$He precludes the use of conventional sample cells that employ a transmission geometry.  Instead, we used a Sk{\"o}ld-Pelizzari reflection cell\cite{Sokol} oriented at $45^\circ$ with respect to the incident beam.  To ensure a correct subtraction of the background signal, the back of the sample cell should scatter no neutrons into the detectors near $\phi = 125^\circ$ when liquid $^3$He is absent.  Boron nitride was cut into a ``sawtooth'' pattern so that neutrons scattering from the back of the sample in direction of the high-angle detectors are absorbed by one of the ``teeth.''  We placed sintered copper powder behind the boron nitride so that the liquid $^3$He could reach thermal equilibrium with the sample cell.

The low temperatures were achieved used a $^3$He sorption cryostat.  Holding the liquid at a constant temperature of 500 mK, we measured $J(Y, Q)$ at pressures $P = (0, 10, 15) \textrm{ bar}$.  This corresponds to number densities $\rho = (0.0163, 0.0197, 0.0216) \textrm{ \AA}^{-3}$.  The experimental data was analyzed using the DAVE software package\cite{DAVE}.

We calculated the effective resolution function $I(Y, Q)$ using a Monte Carlo simulation of the experiment\cite{Azuah}.  The resolution $I(Y, Q)$, shown in Figure \ref{fig:FSE}, is a single Gaussian having a full-width at half-maximum of $0.603 \textrm{ \AA}^{-1}$.  The experimentally observed Compton profile is:
\begin{equation}
J_{EXP}(Y, Q) = \int_{-\infty}^{+\infty} J_{FS}(Y^\prime, Q)I(Y - Y^\prime, Q)dY^\prime.
\end{equation}
Like the FSE function $R(Y, Q)$, the instrumental resolution function $I(Y, Q)$ has the effect of washing out sharp features in $J_{IA}(Y)$.

\section{Lineshape Comparison}

\begin{figure}
	\onefigure[width = \linewidth]{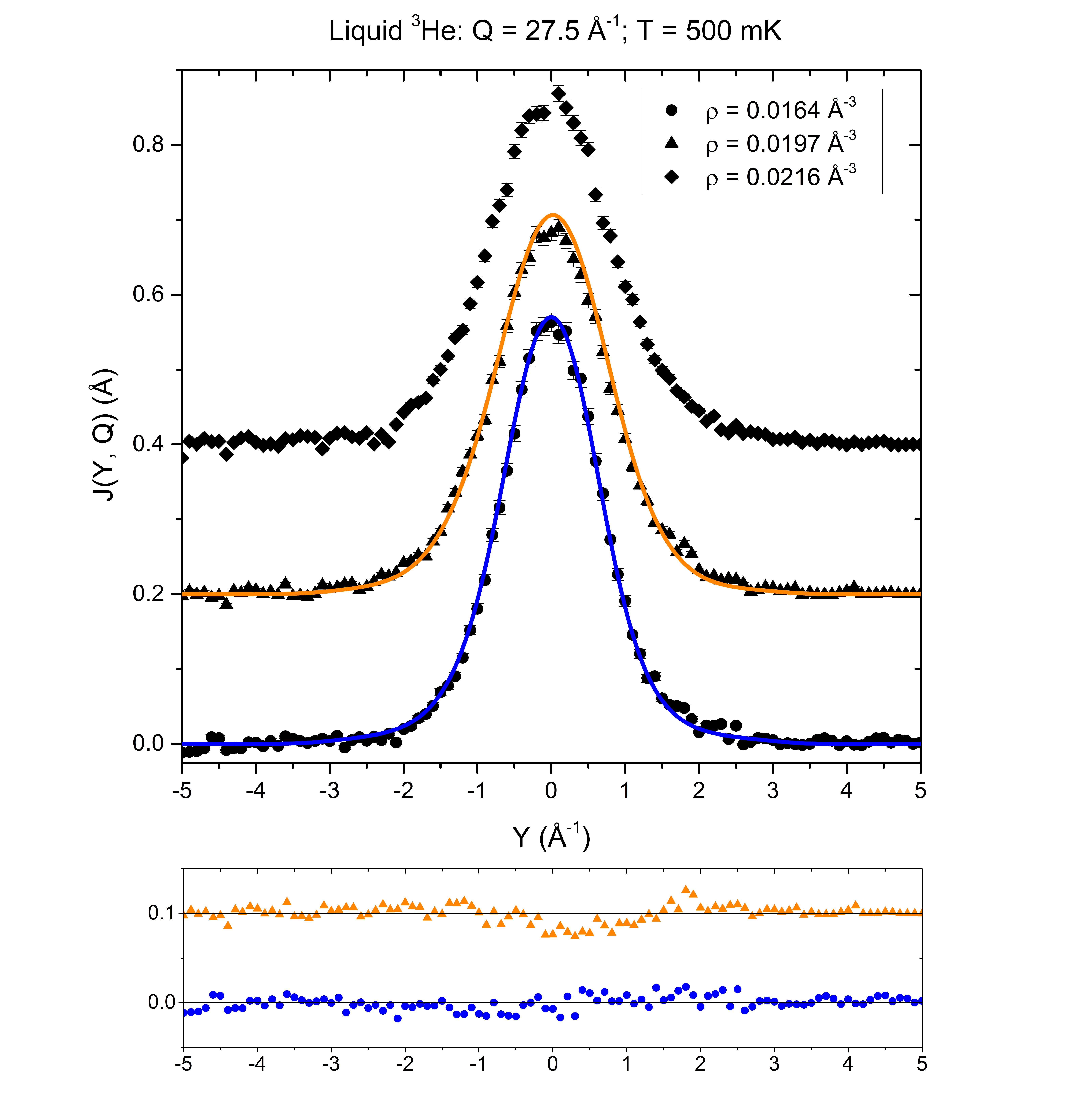}
	\caption{The neutron Compton profile $J(Y, Q)$ of liquid $^3$He at 500 mK at three different densities.  The solid curves are the Diffusion Monte Carlo predictions of Moroni \emph{et al}\cite{Moroni1} when instrumental resolution and final state effect corrections are taken into account.  Difference curves are shown below the main figure.  No calculations are available for comparison with the $\rho = 0.0216 \textrm{ \AA}^{-3}$ data set.  Error bars throughout the text represent one standard deviation.}
\label{fig:DataAndTheory}
\end{figure}

Figure \ref{fig:DataAndTheory} plots the observed Compton profile $J(Y, Q)$ at $Q = 27.5 \textrm{ \AA}^{-1}$ as a function of density.   The scattering consists of a single non-Gaussian peak that becomes broader as the density of the fluid increases.  The increase in width at higher density is a consequence of the Heisenberg uncertainty principle: the zero-point energy of the $^3$He atoms increases as the atoms are more localized.  The same effect has been observed in bulk liquid $^4$He under pressure\cite{Herwig, Diallo}. 

A direct observation of a kink in $J(Y, Q)$ at $Y = k_F$ would constitute conclusive experimental evidence for the existence of a Fermi surface in liquid $^3$He.  No such kinks are observed at any density, even though the temperature of the liquid $^3$He is well below its renormalized Fermi temperature $T_F^* = 1.5$ K.  This sharp feature, readily apparent in the IA prediction $J_{IA}(Y)$ shown in Figure \ref{fig:IA}, is washed out by the combination of instrumental resolution $I(Y, Q)$ and final state effects $R(Y, Q)$ shown in Figure \ref{fig:FSE}.

This situation may be contrasted to X-ray Compton scattering studies of conduction electrons\cite{Cooper, CooperBook}.  The Fermi surface discontinuity is regularly observed in these experiments.  The relative importance of instrumental resolution and final state effects may be estimated by the dimensionless ratio $\Delta Y/k_f$, where $\Delta Y$ is the broadening due to $I(Y, Q)$ and $R(Y, Q)$.  For X-ray Compton scattering experiments, $\Delta Y/k_f$ is on the order of $10^{-1}$ to $10^{-2}$; for neutron Compton scattering studies of liquid $^3$He, $\Delta Y/k_f$ is on the order of $1$.

Theoretical calculations of $n(k)$ may still be checked for their consistency with the scattering data, even if the change in slope at $k_F$ does not appear as a distinct feature in that data.  To make the most stringent possible test, one should compare the entire predicted lineshape for $J(Y, Q)$ with the neutron Compton scattering data.  The solid lines in Figure \ref{fig:DataAndTheory} are obtained when DMC predictions for $J_{IA}(Y)$ are convoluted with the HCPT final state effects $R(Y, Q)$ and instrumental resolution $I(I, Y)$.  As can be seen, there is excellent agreement between theory and experiment with no adjustable parameters. 

The residuals shown in Figure 3 suggest that the predicted lineshape for $J(Y, Q)$ slightly underestimates the scattering intensity near $Y = +1.7 \textrm{ \AA}^{-1}$.  We attribute this small difference to the form of our model FSE function $R(Y, Q)$.  As illustrated in Figure 2, $R(Y, Q)$ has a narrow central peak and oscillatory tails that may be either positive or negative.  The effect of convoluting $J_{IA}(Y)$ with $R(Y, Q)$ is not only to smear the Fermi surface kink, but also to shift intensity around in a way that preserves the $\omega^2$-sum rule.  Specifically, $R(Y, Q)$ depletes the intensity at intermediate, positive $Y$.

\section{Extraction of $\langle E_K\rangle$}
\begin{figure}
	\onefigure[width = \linewidth]{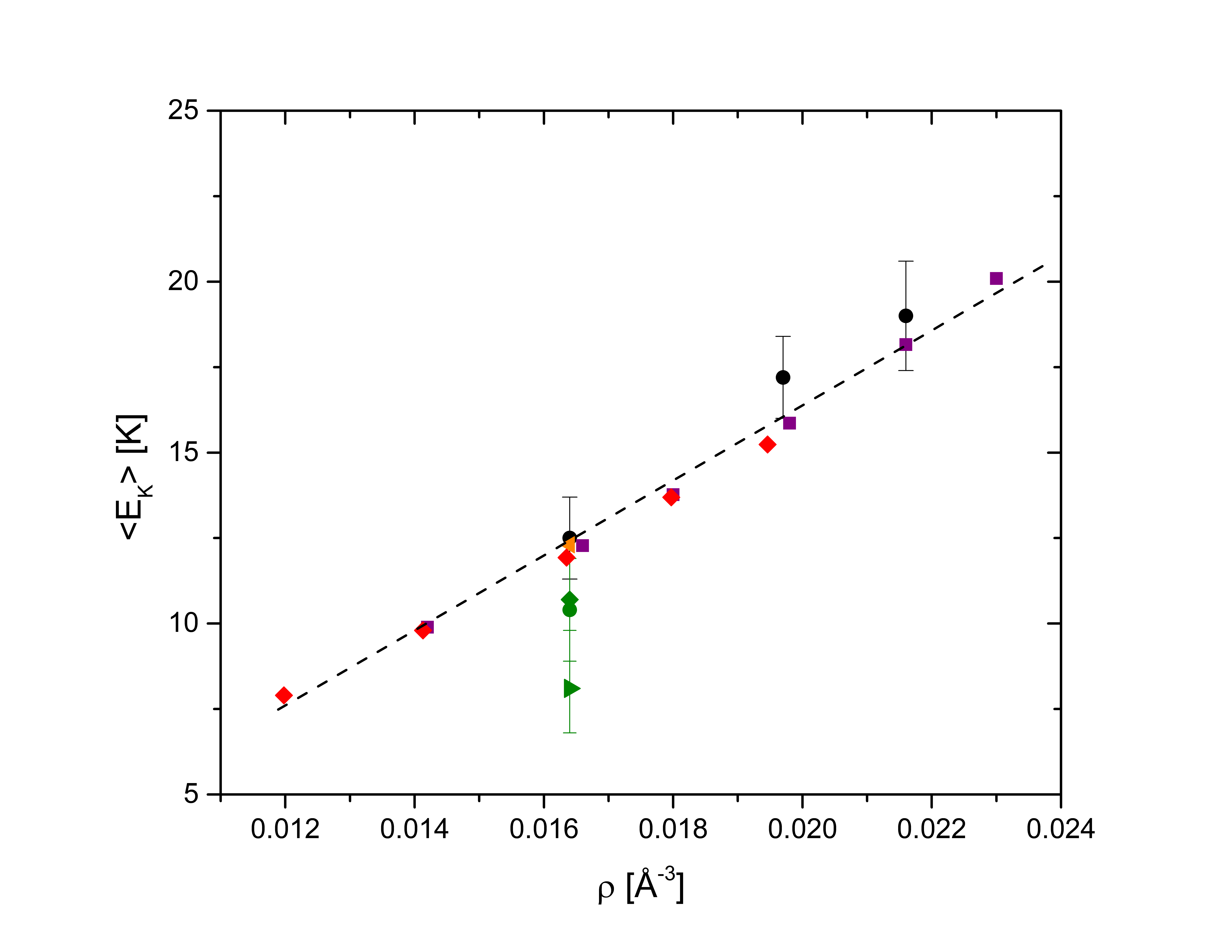}
	\caption{The average kinetic energy $\langle E_K\rangle$ of liquid $^3$He as a function of density: present experimental values (black circles); previous experimental values from Ref\cite{Sokol} (green triangle), Ref\cite{Mook} (green diamond), and Ref\cite{Azuah} (green circle); Diffusion Monte Carlo estimates from Ref\cite{Moroni1} (red diamonds) and Ref\cite{Mazzanti} (orange triangle); and variational estimates from Ref\cite{Manousakis} (purple squares).  The dashed black line is only a guide to the eye and not a fit to the points.}
\label{fig:KE}
\end{figure}

The average kinetic energy $\langle E_K\rangle$ of the $^3$He atoms provides another way to test theoretical calculations of $n(k)$.  According to the $\omega^2$-sum rule, the average kinetic energy $\langle E_K\rangle$ is directly proportional to the second moment of the scattering $J(Y, Q)$.  To determine $\langle E_K\rangle$, one introduces a parameterized model for $J_{IA}(Y)$ and the values of the adjustable parameters are estimated by means of a least squares fit to the scattering data.  The parameterized model $J_{IA}(Y)$ is convoluted with $R(Y, Q)$ and $I(Y, Q)$.

Our preliminary report\cite{Dimeo} of this experiment communicated values for the average kinetic energy $\langle E_K \rangle$ in disagreement with theoretical predictions.  We represented $J_{IA}(Y)$ by means of a single Gaussian when fitting the scattering data.  However, the single Gaussian model is not sufficiently flexible to describe the wings and tails of the scattering, leading to an incorrect determination of $\langle E_K \rangle$.

Here we represent the IA-scattering by means of a phenomenological, non-Gaussian model:
\begin{equation}
J_{IA}^{(P)}(Y) = \sum_{i = 1}^{2} \frac{A_i}{(2\pi\sigma_i^2)^{1/2}}e^{-(Y - Y_0)^2/2\sigma_i^2}.
\label{Pmodel}
\end{equation}
The model scattering function $J_{IA}^{(P)}(Y)$ consists of a sum of two Gaussians, each locked to a common center $Y_0$.  The average kinetic energy is given by:
\begin{equation}
\langle E_K\rangle = \frac{3\hbar^2}{2m}\frac{A_1\sigma_1^2 + A_2\sigma_2^2}{A_1 + A_2}.
\end{equation}

Figure \ref{fig:KE} compares experimental and theoretical estimates for the average kinetic energy $\langle E_K\rangle$.  We applied the $\omega^2$-sum rule to the DMC predictions in Ref\cite{Moroni1} to obtain the red diamonds shown in the figure.  All previous experiments at saturated vapor pressure disagree with theoretical predictions.  In contrast, the present results are in good agreement with theoretical predictions at all densities.  For pressures of 0, 10, 15 bar, we estimate $\langle E_K \rangle$ is $12.5 \textrm{ K} \pm 1.2 \textrm{ K}$, $17.2 \textrm{ K} \pm 1.2 \textrm{ K}$, and $19.0 \textrm{ K} \pm 1.6 \textrm{ K}$, respectively.

\section{Alternative Models}
We have shown that the conventional Fermi liquid picture of $^3$He is consistent with our neutron Compton scattering data.  The question naturally arises whether or not the experimental data by itself proves that a Fermi surface discontinuity exists in liquid $^3$He.  Bouchaud and Lhuillier developed an alternative model in which $^3$He atoms form BCS-like dimers and no sharp Fermi surface exists in $n(k)$\cite{BL, BL2}.  Unfortunately, numerical calcuations of $n(k)$ for the Bouchaud-Lhuillier scenerio are not well developed enough in the literature to permit a detailed comparison with our results.  

The measured scattering from liquid $^3$He is consistent with many possible forms for the momentum distribution $n(k)$, including models without a Fermi surface discontinuity or an exponential tail.  Both of these characteristics are absent in the phenomenological model $J_{IA}^{(P)}(Y)$, which represents $n(k)$ as a sum of two Gaussians.  This implies that the problem of inverting the scattering data $J(Y, Q)$ to a unique momentum distribution $n(k)$ is ill-posed. 

This is analogous to neutron Compton scattering studies of superfluid $^4$He\cite{Sosnick, Snow, GriffinBEC, Azuah4He, Glyde4He}.  Here the Bose-Einstein condensate contributes a $\delta$-function singularity at $Y = 0$ to the IA-scattering $J_{IA}(Y)$.  While this $\delta$-function is not present in the observed scattering $J(Y, Q)$, there is an increase in intensity at small $Y$ when the liquid $^4$He is cooled from the normal fluid into the superfluid phase.  This increase in scattering is consistent with the existence of a condensate peak broadened by finite instrumental resolution and final state effects.  However, it is also consistent with other possible forms for $n(k)$, such as a sum of two Gaussians.  Sivia and Silver showed that neutron Compton profile $J(Y, Q)$ of superfluid $^4$He cannot be uniquely inverted to $n(k)$ even for data with very high statistical precision\cite{SiviaSilver}.

\section{Conclusion}
In this paper, we have presented a neutron Compton scattering study of normal liquid $^3$He at 500 mK under 0, 10, and 15 bars of applied pressure.  We directly compared the observed scattering to \emph{ab initio} predictions and excellent agreement is obtained at all pressures.  The average atomic kinetic energy $\langle E_K\rangle$ is also in good agreement with theoretical expectations.  The new results resolve a long-standing inconsistency between theoretical calculations of $\langle E_K\rangle$ and neutron Compton scattering experiments.

\acknowledgments
This report was prepared by Indiana University under award 70NANB10H255 from the National Institute of Standards and Technology.  Matthew S. Bryan acknowledges support under NSF grant DGE-1069091.  Experiments at the ISIS Pulsed Neutron and Muon Source were supported by a beamtime allocation from the Science and Technology Facilities Council.  The statements, findings, conclusions, and recommendations are those of the authors and do not necessarily reflect the views of the National Institute of Standards and Technology or the U.S. Department of Commerce.   The authors gratefully acknowledge the assistance of K. Guckelsberger with the neutron scattering experiment.

\end{document}